# Real-time Prediction of Bitcoin Bubble Crashes

Min Shu[1, 2, *], Wei Zhu[1, 2]

[1] Department of Applied Mathematics & Statistics, Stony Brook University, Stony Brook, NY, USA
[2] Center of Excellence in Wireless & Information Technology, Stony Brook University, Stony Brook, NY, USA

**Abstract**

In the past decade, Bitcoin as an emerging asset class has gained widespread public attention because of their extraordinary returns in phases of extreme price growth and their unpredictable massive crashes. We apply the log-periodic power law singularity (LPPLS) confidence indicator as a diagnostic tool for identifying bubbles using the daily data on Bitcoin price in the past two years. We find that the LPPLS confidence indicator based on the daily Bitcoin price data fails to provide effective warnings for detecting the bubbles when the Bitcoin price suffers from a large fluctuation in a short time, especially for positive bubbles. In order to diagnose the existence of bubbles and accurately predict the bubble crashes in the cryptocurrency market, this study proposes an adaptive multilevel time series detection methodology based on the LPPLS model and finer (than daily) timescale for the Bitcoin price data. We adopt two levels of time series, 1 hour and 30 minutes, to demonstrate the adaptive multilevel time series detection methodology. The results show that the LPPLS confidence indicator based on this new method is an outstanding instrument to effectively detect the bubbles and accurately forecast the bubble crashes, even if a bubble exists in a short time. In addition, we discover that the short-term LPPLS confidence indicator highly sensitive to the extreme fluctuations of Bitcoin price can provide some useful insights into the bubble status on a shorter time scale - on a day to week scale, and the long-term LPPLS confidence indicator has a stable performance in terms of effectively monitoring the bubble status on a longer time scale - on a week to month scale. The adaptive multilevel time series detection methodology can provide real-time detection of bubbles and advanced forecast of crashes to warn of the imminent risk in not only the cryptocurrency market but also other financial markets.

Keywords: Cryptocurrency; Bitcoin; Bitcoin bubble crash; Log-periodic power law singularity (LPPLS); Market crashes; Adaptive Multilevel Time Series Detection Methodology

---

*Corresponding author at: Department of Applied Mathematics & Statistics, Physics A149, Stony Brook University, Stony Brook, NY 11794, USA.
*E-mail address:* min.shu@stonybrook.edu (M. Shu), wei.zhu@stonybrook.edu (W. Zhu)

# 1. Introduction

In the past decade, cryptocurrencies as digital coins have become well known because of their extraordinary return potential in phases of extreme price growth and their unpredictable massive crashes. A cryptocurrency is designed to act as a medium of exchange, using powerful encryption techniques to protect financial transactions, control the creation of other units, and verify the transfer of assets. As opposed to centralized digital currency and central banking systems, cryptocurrencies are decentralized through a distributed ledger technology known as blockchain that acts as a public financial transaction database [1]. Due to the failure of governments and central banks in the 2008 global financial crisis and the 2010-2013 European Sovereign Debt Crisis, the cryptocurrencies have attracted soaring attentions from various economic actors. Since cryptocurrency as an asset class is still in the nascent stages, the price of cryptocurrency has some remarkable fluctuations. As shown in Figure 1 (https://coinmarketcap.com), the total market capitalization of cryptocurrencies had climbed to the historical peak of $835.69 billion on January 7, 2018, and then crashed to $283.95 billion on February 6, 2018, dropping by at least 66% over merely 30 days.

The first decentralized cryptocurrency, Bitcoin, was introduced by pseudonymous Satoshi Nakamoto in 2008 and then released as an open-source software in 2009 based on the proof-of-work scheme of SHA-256 cryptographic hash function [2]. After the release of Bitcoin, over 4,000 cryptocurrencies have since been created. By the end of December 2018, the total number of Blockchain wallet users worldwide reaches nearly 32 million [3]. Currently, there are almost 7.1 million active bitcoin users and five percent of Americans are estimated to hold Bitcoin [4]. Since its inception in 2009, Bitcoin has managed to keep its leading position in the cryptocurrency market. On April 3rd, 2019, the total market capitalization of Bitcoin as shown in Figure 1 is around $92.8 billion, consisting of 50.2% of the total market capitalizations of cryptocurrencies at $184.7 billion.

Bitcoin has been characterized by sharp upward and downward price movements with large volatility. Figure 2 shows the evolution of the price trajectories of the Bitcoin from September 13, 2011 to April 7, 2019 (https://bitcoincharts.com/). The logarithm of Bitcoin price in US Dollars is plotted as a function of time so that an exponential growth with constant growth rate can be represented by a straight line. The dashed straight line is the best linear fit of the logarithm of Bitcoin price from 2011 to 2019, that represents the long-term behavior of Bitcoin price trajectory. The compound annual growth rate of the Bitcoin price is determined as 170.5% per year. The evolution of the price trajectories of the Bitcoin shows that the price of the long-term exponential growth has to suffer a succession of bubbles and crashes.

Its extraordinary return potential and its technological and economic prospects have intensified the economics and finance debates on Bitcoin from the scientific and investment communities. Many pundits have claimed that the value of Bitcoin as a fraud will eventually go to zero. Bitcoin has been criticized for its unstable price, potential economic bubble, illegal transaction usage, high energy consumption, lofty and variable transactions costs, poor security and fraud at cryptocurrency exchanges, vulnerability to debasement, and the influence of miners. Hanley [5] pointed out the false claims of Bitcoin and concluded that bitcoin is not credible and has no intrinsic value to support the market valuation. Yermack [6] argued that Bitcoin has no fundamental value and largely fails to satisfy the criteria of acting as a medium of exchange, a store of value, and a unit of account, and altogether behaving more like a speculative investment than a currency. In contrast, many experts have argued that Bitcoin will have enormous growth and be wildly adopted in the future, and eventually revolutionize the conventional transaction and banking system. Popper [7] treated the Bitcoin as the digital gold. The Nobel Prize-winning economist Robert Shiller pointed out the exceptionally ambiguous value of Bitcoin and, at the 2018 Davos World, stated that "Bitcoin could be here for 100 years but it's more likely to totally collapse" and, "you just put an upper bound on [Bitcoin] with the value of the world's money supply. But that upper bound is awfully big. So it can be anywhere between zero and there." [8].

There is an emerging academic literature on the statistical properties and predictions of the Bitcoin financial time series. Kristoufek [9] observed the bidirectional relationship between the Bitcoin price and the web search queries on Google Trends and Wikipedia. Garcia, Tessone, Mavrodiev and Perony [10] detected the two positive feedback loops: the social cycle between search volume, volume of word-of-mouth communication and price, and the user adoption cycle between search volume, number of new users and price. Glaser, Zimmermann, Haferkorn, Weber and Siering [11] investigated the links among Bitcoin price, Blochain transaction and netwok volume, and search query data, and found that Bitcoin users are primarily interested in trading Bitcoin as a speculative investment rather than an alternative transaction system. Donier and Bouchaud [12] predicted the size of potential crashes by analyzing the liquidity of the Bitcoin market based on the order book data. Bouri, Molnár, Azzi, Roubaud and Hagfors [13] investigated the hedge and safe haven properties against other major financial assets based on the dynamic conditional correlation model, and their results showed that Bitcoin can only be useful as an effective diversifier, but not as a hedging tool. Balcilar, Bouri, Gupta and Roubaud [14] detected the nonlinear causal relationships between Bitcoin returns and trading volume under normal market conditions by conducting the causality-in-quantiles tests. Bariviera [15] studied the long memory of the return and volatility on Bitcoin using the Hurst exponent by the Rescaled range analysis method and the Detrended Fluctuation Analysis method, and concluded that the long memory is exhibited in daily volatility, while suffering from regime switch in the daily return. Begušić, Kostanjčar, Stanley and Podobnik [16] found that the tail of Bitcoin returns cumulative distribution does not follow the inverse cubic law which is a well-known characteristic of financial asset returns and the value of the scaling exponent is in the range from 2.0 to 2.5. McNally, Roche and Caton [17] implemented a Bayesian optimal recurrent neural network (RNN) and a long short term memory (LSTM) network to predict the price of Bitcoin. Wheatley, Sornette, Huber, Reppen and Gantner [18] evaluated the fundamental value of Bitcoin based on the network properties and predicted the bubbles using the Market-to-Metcalfe Ratio. Gerlach, Demos and Sornette [19] presented a peak detection method based on the Epsilon Drawdown method and listed the possible socioeconomic causes of the Bitcoin bubbles, as well as detailed predictive analysis of the bubbles using k-mean clustering method.

Due to the sharp change and large volatility of cryptocurrency price, the conventional financial bubble detection methods do not work well to provide an effective identification of bubbles in the cryptocurrency market. This study has proposed the adaptive multilevel time series detection methodology based on the Log Periodic Power Law Singularity (LPPLS) model to diagnose bubbles and forecast crashes in the cryptocurrency market. In LPPLS model, the bubbles are considered as the result of unsustainable (faster than exponential) growth to achieve an infinite return in finite time (a finite time singularity), forcing a correction (change of regime) of an asset price in a real world. The LPPLS model is built upon the theory of rational expectation and combines two well documented empirical and phenomenological features of bubbles: the transient super-exponential growth and the accelerating log-periodic volatility fluctuations. The first characteristic is caused by positive feedback mechanism in the valuation of assets created by imitation and herding behavior of noise traders and of boundedly rational agent (Johansen et al., 1999). This characteristic can be modeled by a hyperbolic power law with a singularity in finite time, so that a crash or correction is eventually necessitated before the price approaches to an infinite value at a singularity time point endogenously. The second characteristic results from the spirals of competing expectations of higher returns (bullish) and an impending crash (bearish) (Johansen et al., 1999). The log-period fluctuation is ubiquitous in complex systems with hierarchical structures resulting from the interplay between the restoring mechanism and the nonlinear growth rate [20]. Figure 2 unveils these two characteristics of the Bitcoin bubbles. A succession of price run-ups characterized by increased growth rates can be observed in Figure 2, which is reflected visually by transient processes characterized by strong upward curvature of the price trajectory. Such an upward curvature in a linear-log plot provides a first visual diagnostic of a transient super-exponential.

In recent years, the LPPLS model has been studied by many researchers for bubble detection. Filimonov and Sornette [21] transformed the LPPLS formulation and condensed the number of nonlinear parameters in the function from four to three to reduce the calibration complexity. Lin, Ren and Sornette [22] developed a self-consistent model for explosive financial bubbles which combines a mean-reverting volatility process and a stochastic conditional return. Sornette, Demos, Zhang, Cauwels, Filimonov and Zhang [23] proposed the LPPLS Confidence indicator and the LPPLS Trust indicator to evaluate the performance of the real-time prediction of bubble crash in 2015 Shanghai stock market. Filimonov, Demos and Sornette [24] calibrated the LPPLS model by applying the modified profile likelihood inference method and obtained the interval estimation for the critical time. The LPPLS model provided a flexible framework to detect financial bubbles through analyzing the time series of asset prices.

The rest of this paper is organized as follows: Section 2 presents our methodology; Section 3 discusses the analysis and the results; and, finally, Section 4 concludes this study.

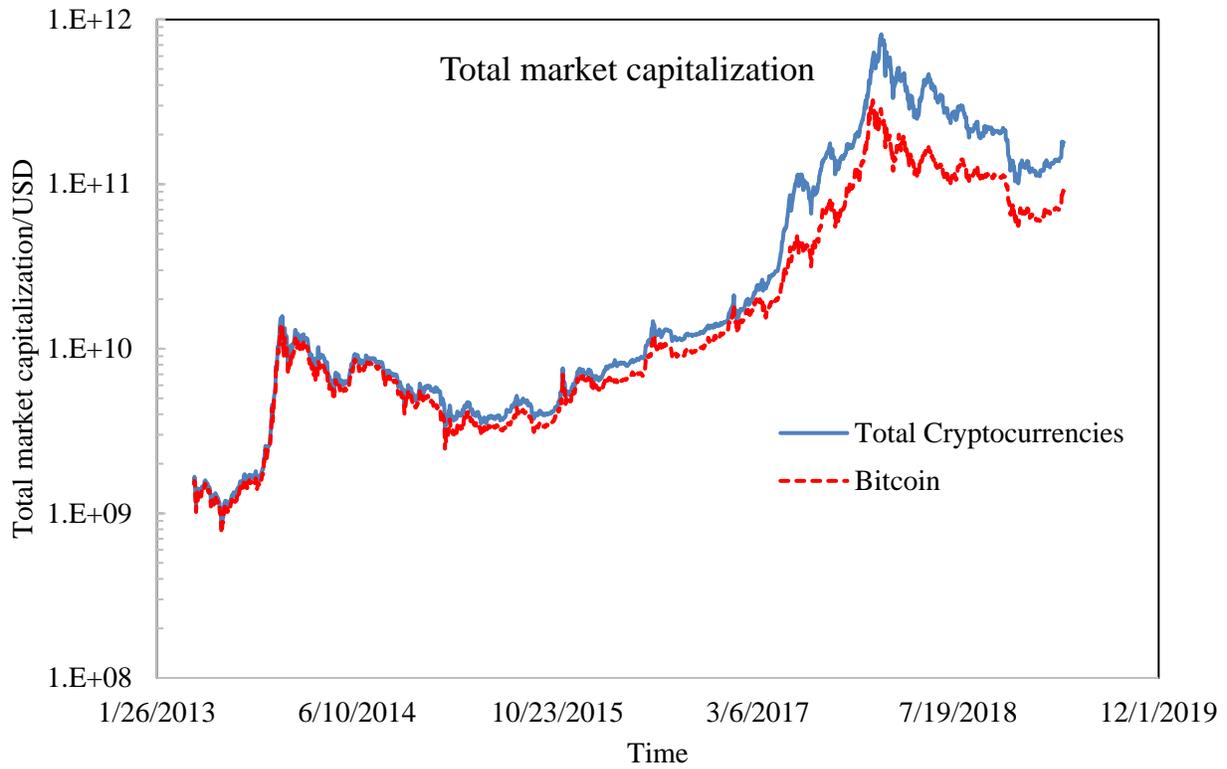

Figure 1. Evolution of the total market capitalization of cryptocurrencies and Bitcoin from April 2013 to April 2019

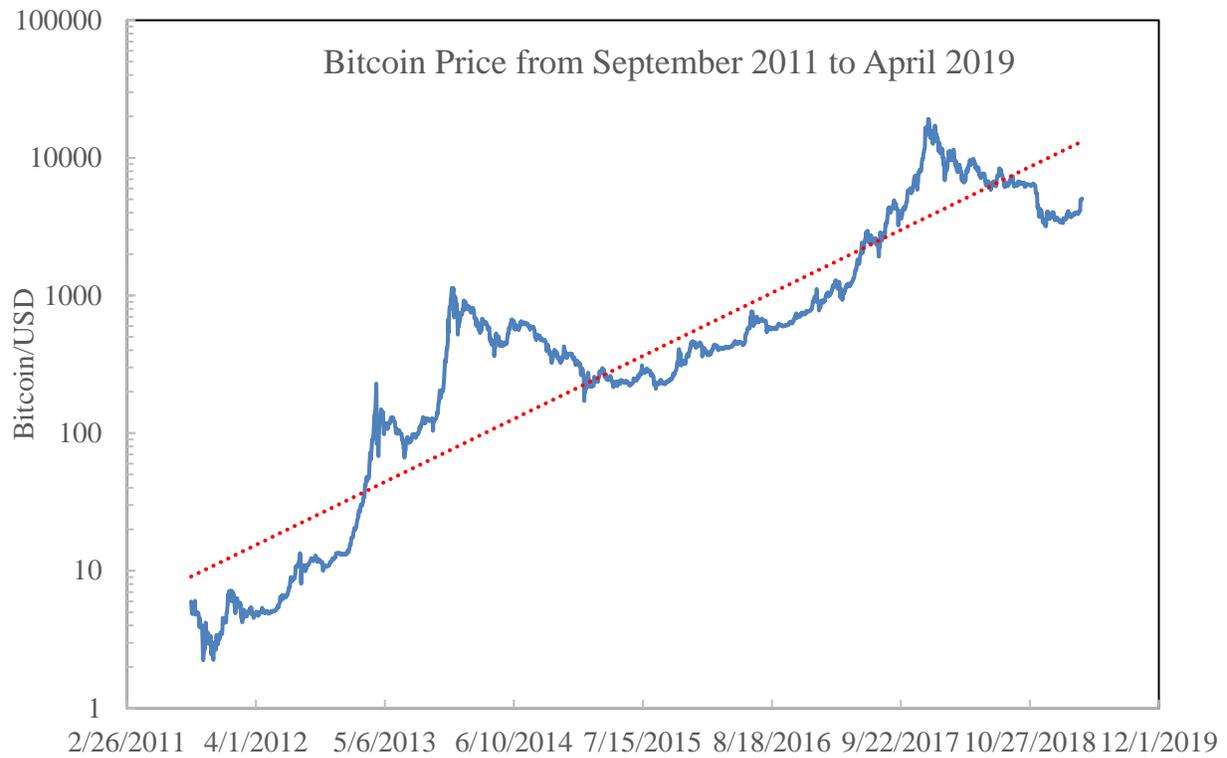

Figure 2. Evolution of the price trajectories of the Bitcoin

## 2. Methodology

*2.1 The Log-Periodic Power Law Singularity (LPPLS) Model and Calibration*

In a bubble regime, the observed price trajectory of a given asset decouples from its intrinsic fundamental value and performs two typical characteristics: the transient super-exponential growth and the accelerating log-periodic volatility fluctuations. The LPPLS model combines these two typical bubble characteristics to model the bubble as a process of super-exponential power law growth punctuated by short-lived corrections organized according to the symmetry of discrete scale invariance [25]. The LPPLS model is based on (i) the economic theory of rational expectation bubbles, (ii) behavioral finance on imitation and herding of traders, and (iii) the mathematical and statistical physics of bifurcations and phase transitions [26].

The LPPLS model, originally called as the Johansen-Leoit-Sornette (JLS) model firstly proposed by Sornette, Johansen and Bouchaud [27], assumes that the observed asset price $p(t)$ can be described as:

$$\frac{dp}{p} = \mu(t)dt + \sigma(t)dW - \kappa dj \tag{1}$$

where $\mu(t)$ is the expected return, $\sigma(t)$ is the volatility, $dW$ is the increment of a standard Wiener process with zero mean and unit variance, $\kappa$ is the loss amplitude when a crash occurs, and $dj$ represents a discontinuous jump with the value of 0 before the crash and 1 after the crash. The dynamics of the jumps $dj$ is governed by a crash hazard rate $h(t)$. $h(t)$ represents the crash probability at a specified time $t$, thus $h(t)dt$ is the bubble crash probability between $t$ and $t + dt$ conditional on the fact that it has not yet happened. The expectation of jump $dj$ can be determined as:

$$E_t[dj] = 1 \times h(t)\mathrm{d}t + 0 \times (1 - h(t)\mathrm{d}t) = h(t)\mathrm{d}t \tag{2}$$

In the network structure underlying the LPPLS model, two types of agents in a market are considered: one group is formed by traders with rational expectations and the other group consists of noise traders whose decisions to buy, sell, or hold are irrational and erratic. The noise trader is susceptible to show imitation and herding behavior and destabilize the asset price. Johansen, Ledoit and Sornette [28] proposed that the behavior of the agent network can be incorporated by quantifying the crash hazard rate $h(t)$ in the following form:

$$h(t) = \alpha(t_c - t)^{m-1}(1 + \beta cos(\omega \ln(t_c - t) - \phi)) \tag{3}$$

where $\alpha, \beta, m, \omega,$ and $\phi$ are the parameters. The power law singularity $(t_c - t)^{m-1}$ embodies the mechanism of positive feedback associated with the herding behavior of the noise traders, leading to the formation of bubbles. The power law singularity reaches the singularity at the time $t$ is equal to the critical time $t_c$. The log periodic function $\cos(\omega \ln(t_c - t) - \phi)$ takes into account the existence of a possible hierarchical cascade of panic acceleration punctuating the growth of the bubble, resulting either from a preexisting hierarchy in noise trader sizes [29] and/or from the interplay between market price impact inertia and nonlinear fundamental value investing [20].

Under the assumption of no arbitrage condition and rational expectations, the conditional expectation of the price dynamics $E_t[dp] = 0$ as the price process satisfies the martingale condition. The expectation of Equation (1) can be determined as:

$$E_t[dp] = \mu(t)p(t)dt + \sigma(t)p(t)E_t(dW) - \kappa p(t)E_t(dj) \tag{4}$$

Since $E_t[dW] = 0$, and $E_t[dj] = h(t)dt$, Equation (4) yields:

$$\mu(t) = \kappa h(t) \tag{5}$$

Equation (5) indicates the return $\mu(t)$ is controlled by the bubble crash risk quantified by the crash hazard rate $h(t)$.

Conditioning on the fact that no crash occurs, Equation (1) can be simplified as:

$$\frac{dp}{p} = \mu(t)dt + \sigma(t)dW - \kappa \times 0 = \kappa h(t)dt + \sigma(t)dW \tag{6}$$

The conditional expectation of Equation (6) leads to:

$$E_t[\frac{dp}{p}] = \kappa h(t)dt \tag{7}$$

Solving Equation (7) by substituting Equation (3) under the condition that no crash has yet occurred leads to the simple mathematical formulation of the LPPLS for the expected value of a log-price [30]:

$$\text{LPPLS}(t) \equiv \ln E[p(t)] = A + B(t_c - t)^m + C(t_c - t)^m \cos[\omega \ln(t_c - t) - \phi] \tag{8}$$

Here $A > 0$ is the expected value of the $\ln p(t)$ at the critical time $t_c$. $B = -k\alpha/m$ and $B < 0$ ($B > 0$) ensures that the price is indeed growing (decreasing) super-exponentially as time goes towards $t_c$. $C = -k\alpha\beta/\sqrt{m^2 + \omega^2}$ is the proportional magnitude of the oscillations around the power law singular growth. $0 < m < 1$ is the exponent of the power law growth. The first condition of $m > 0$ ensures that the price remains finite at $t_c$, while $m < 1$ makes sure a singularity exist, that is, the expected log-price diverges at $t_c$ for $m < 1$. The critical time $t_c$ is the theoretical termination time of a financial bubble. A bubble could be terminated by a large crash or a change of the average growth rate (regime change), so that the super-exponential growth rate of price changes to an exponential or lower growth with the end of the accelerating oscillations. $\omega$ is the angular log-frequency of the oscillation and $0 < \phi < 2\pi$ is a phase parameter.

The remarkable characteristic of the super-exponential growth of the bubble can be described by the power law singular component $A + B(t_c - t)^m$, which embodies the positive feedback mechanism of a bubble development. The characteristic of accelerating oscillations of the bubble is captured by the component $C(t_c - t)^m \cos[\omega \ln(t_c - t) + \phi]$, which represents the tension and competition between the value investors and the noise traders resulting in the deviation of the market price around the super-exponential growth in the form of oscillations that are periodic in the logarithm of the time to $t_c$. The term $C(t_c - t)^m$ expresses the fact that the amplitude of the accelerating oscillation is falling to zero at $t_c$. The term $\omega \ln(t_c - t)$ represents the local frequency of the log-periodic oscillations is accelerating to infinite at $t_c$.

The term $C \cos[\omega \ln(t_c - t) - \varphi]$ in Equation (8) can be expanded to replace the two parameters $C$ and $\varphi$ by two linear parameters $C_1 = C\cos\phi$ and $C_2 = C\sin\phi$. The transformed LPPLS formula is written as [21]:

$$\text{LPPLS}(t) \equiv E_t[\ln p(t)] = A + B(t_c - t)^m + C_1(t_c - t)^m \cos[\omega \ln(t_c - t)] \\ + C_2(t_c - t)^m \sin[\omega \ln(t_c - t)] \quad (9)$$

The reformed LPPLS model has 3 nonlinear parameters $\{t_c, m, \omega\}$ and 4 linear parameters $\{A, B, C_1, C_2\}$, and the phase $\phi$ is contained by $C_1$ and $C_2$. Using the $L^2$ norm, the sum of squares of residuals of Equation (9) can be described as:

$$F(t_c, m, \omega, A, B, C_1, C_2) = \sum_{i=1}^{N} [\ln p(t_i) - A - B(t_c - t_i)^m - C_1(t_c - \tau_i)^m \cos(\omega \ln(t_c - t_i)) \\ - C_2(t_c - t_i)^m \sin(\omega \ln(t_c - t_i))]^2 \quad (10)$$

Slaving the 4 linear parameters $\{A, B, C_1, C_2\}$ to the remaining 3 nonlinear parameters $\{t_c, m, \omega\}$ yields the following cost function $\chi^2(t_c, m, \omega)$:

$$\chi^2(t_c, m, \omega) := F_1(t_c, m, \omega) = \min_{\{A, B, C_1, C_2\}} F(t_c, m, \omega, A, B, C_1, C_2) = F(t_c, m, \omega, \hat{A}, \hat{B}, \hat{C}_1, \hat{C}_2) \quad (11)$$

where the hat symbol ^ indicates estimated parameters. The 4 linear parameters can be estimated by solving the optimization problem

$$\{\hat{A}, \hat{B}, \hat{C}_1, \hat{C}_2\} = \arg \min_{\{A, B, C_1, C_2\}} F(t_c, m, \omega, A, B, C_1, C_2) \quad (12)$$

which can be done analytically by solving the following matrix equations

$$\begin{pmatrix} N & \sum f_i & \sum g_i & \sum h_i \\ \sum f_i & \sum f^2_i & \sum f_i g_i & \sum f_i h_i \\ \sum g_i & \sum f_i g_i & \sum g^2_i & \sum h_i g_i \\ \sum h_i & \sum f_i h_i & \sum g_i h_i & \sum h^2_i \end{pmatrix} \begin{pmatrix} \hat{A} \\ \hat{B} \\ \hat{C}_1 \\ \hat{C}_2 \end{pmatrix} = \begin{pmatrix} \sum \ln p_i \\ \sum f_i \ln p_i \\ \sum g_i \ln p_i \\ \sum h_i \ln p_i \end{pmatrix} \quad (13)$$

where $f_i = (t_c - t_i)^m$, $g_i = (t_c - t_i)^m \cos(\omega \ln(t_c - t_i))$, and $h_i = (t_c - t_i)^m \sin(\omega \ln(t_c - t_i))$.

The 3 nonlinear parameters $\{t_c, m, \omega\}$ can be determined by solving the following nonlinear optimization problem:

$$\{\hat{t}_c, \hat{m}, \hat{\omega}\} = arg \min_{\{t_c, m, \omega\}} F_1(t_c, m, \omega) \tag{14}$$

The LPPLS model is calibrated on the time series of price using the Ordinary Least Squares method, providing estimations of all parameters $\{t_c, m, \omega, A, B, C_1, C_2\}$. In this study, the covariance matrix adaptation evolution strategy (CMA-ES) proposed by Hansen, Ostermeier and Gawelczyk [31] is adopted to search for the best estimation of the three nonlinear parameters $\{t_c, m, \omega\}$ by minimizing the sum of residuals between the fitted LPPLS model and the observed price time series. The CMA-ES rates among the most successful evolutionary algorithms for real-valued single-objective optimization and is typically applied to difficult nonlinear non-convex black-box optimization problems in continuous domain and search space dimensions between three and a hundred. Parallel computing is adopted to expedite the fitting process drastically.

*2.2 LPPLS Confidence Indicator*

The LPPLS confidence indicator proposed by Sornette, Demos, Zhang, Cauwels, Filimonov and Zhang [23] is defined as the fraction of fitting windows in which the LPPLS calibrations satisfy the specified filter conditions. It is used to measure sensitivity of observed bubble pattern to the selection of the start time $t_1$ in fitting windows. The larger LPPLS confidence indicator, the more reliable of the LPPLS bubble pattern. A small value of the indicator signals a possible fragility because the LPPLS bubble patterns are presented in few fitting time series of price. A LPPLS confidence indicator for a specified data point $t_2$ can be calculated in the five steps: (1) shrink the time window by moving the start time $t_1$ toward the endpoint $t_2$ with a specific step of $dt$ to create a group of time series of price, (2) determine the search space of the fitting parameters, (3) calibrate the LPPLS model for each fitting time window, (4) filtrate the calibration results, and (5) calculate the LPPLS confidence indicator from dividing the number of time windows satisfying the specified filter conditions by the total number of the fitting windows. Since the LPPLS confidence indicator is estimated only based on data prior to $t_2$, the value of the LPPLS confidence indicator is causal. The time development of the bubble signal can be detected by the LPPLS confidence indicators for a series of varying $t_2$.

In this study, a group of time series of price for a specified endpoint $t_2$ which is corresponding to a fictitious "present" up to when the data is recorded is created by shrinking the time window ($t_1$, $t_2$) of length $dt = t_2 - t_1 + 1$ decreasing from 650 data points to 30 data points in the step of 5 data points. Thus, 125 fitting windows are determined for each $t_2$. To address the slackness of the model, the search space is limited to:

$$m \in [0,1], \omega \in [1,50], t_c \in \left[t_2, t_2 + \frac{t_2 - t_1}{3}\right], \frac{m|B|}{\omega\sqrt{C_1^2 + C_2^2}} \geq 1 \quad (15)$$

The condition $t_c \in [t_2, t_2 + (t_2 - t_1)/3]$ ensures that the predicted $t_c$ should be after the endpoint $t_2$, and not be too far away from the $t_2$ since the predictive capacity degrades far beyond $t_2$ [32]. The Damping parameter $m|B|/\left(\omega\sqrt{C_1^2 + C_2^2}\right) \geq 1$ expresses the crash hazard rate $h(t)$ is non-negative by definition [33].

After calibrating the LPPLS models, the solutions should be filtered under the stricter conditions:

$$m \in [0.01, 0.99], \omega \in [2, 25], t_c \in \left[t_2, t_2 + \frac{t_2 - t_1}{5}\right], \frac{\omega}{2}\ln\left(\frac{t_c - t_1}{t_c - t_2}\right) \geq 2.5,$$

$$\max\left(\frac{|\hat{p}_t - p_t|}{p_t}\right) \leq 0.15, \ p_{lomb} \leq \alpha_{sign}, \ \ln(\hat{p}_t) - \ln(p_t) \sim AR(1) \quad (16)$$

The filter conditions are derived from the empirical evidence gathered in investigations of previous bubbles [23, 32] and are the stylized features of LPPLS model. The condition for the number of oscillations (half-periods) of the log-periodic component $(\omega/\pi)\ln[(t_c - t_1)/(t_c - t_2)] \geq 2.5$ is adopted to distinguish a genuine log-periodic signal from one that could be generated by noise [34].

The maximum relative error $\max(|\hat{p}_t - p_t|/p_t) \leq 0.15$ is used to ensure that the fitted price of an asset $\hat{p}_t$ should be not too far from the actual asset price $p_t$. The condition $P_{lomb} \leq \alpha_{sig}$ ensures the logarithm-periodic oscillations in fitting the logarithm of prices to the LPPLS model by applying the Lomb spectral analysis for the time series of detrended residual $r(t) = (t_c - t)^{-m}(\ln[p(t)] - A - B(t_c - t)^m)$ [35]. The probability that the maximum peak occurred by chance $P_{lomb}$ is less than the specified significant level $\alpha_{sig}$, indicating the existence the logarithm-periodic oscillations in the fitting LPPLS model. The $\ln(\hat{p}_t) - \ln(p_t) \sim AR(1)$ condition expresses that the LPPLS fitting residuals can be modeled by a mean-reversal Ornstein-Uhlenbeck (O-U) process when the logarithmic price in the bubble regime is attributed to a deterministic LPPLS component [22]. Since the test for the O-U property of LPPLS fitting residuals can be translated into an AR(1) test for the corresponding residuals, both the Phillips-Perron unit-root test and the Dickey-Fuller unit-root test are used to check the O-U property of LPPLS fitting residuals. In this study, the 5% significant level is applied in the tests. Only the calibration results satisfying filter conditions given in Equation (16) are considered valid while the rest discarded.

*2.3 Adaptive Multilevel Time Series Detection Methodology.*

Compared to conventional stock market, the cryptocurrency's price has a larger volatility and a sharper fluctuation in a shorter time. It is necessary to use time series of price with a smaller time interval to capture the accurate features of price fluctuation. In order to effectively diagnose the existence of bubble and accurately predict its crash in the cryptocurrency market, we propose a novel adaptive multilevel time series detection methodology and framework based on the LPPLS model. The flow-chart of the adaptive multilevel time series detection analysis framework is presented in Figure 1 below.

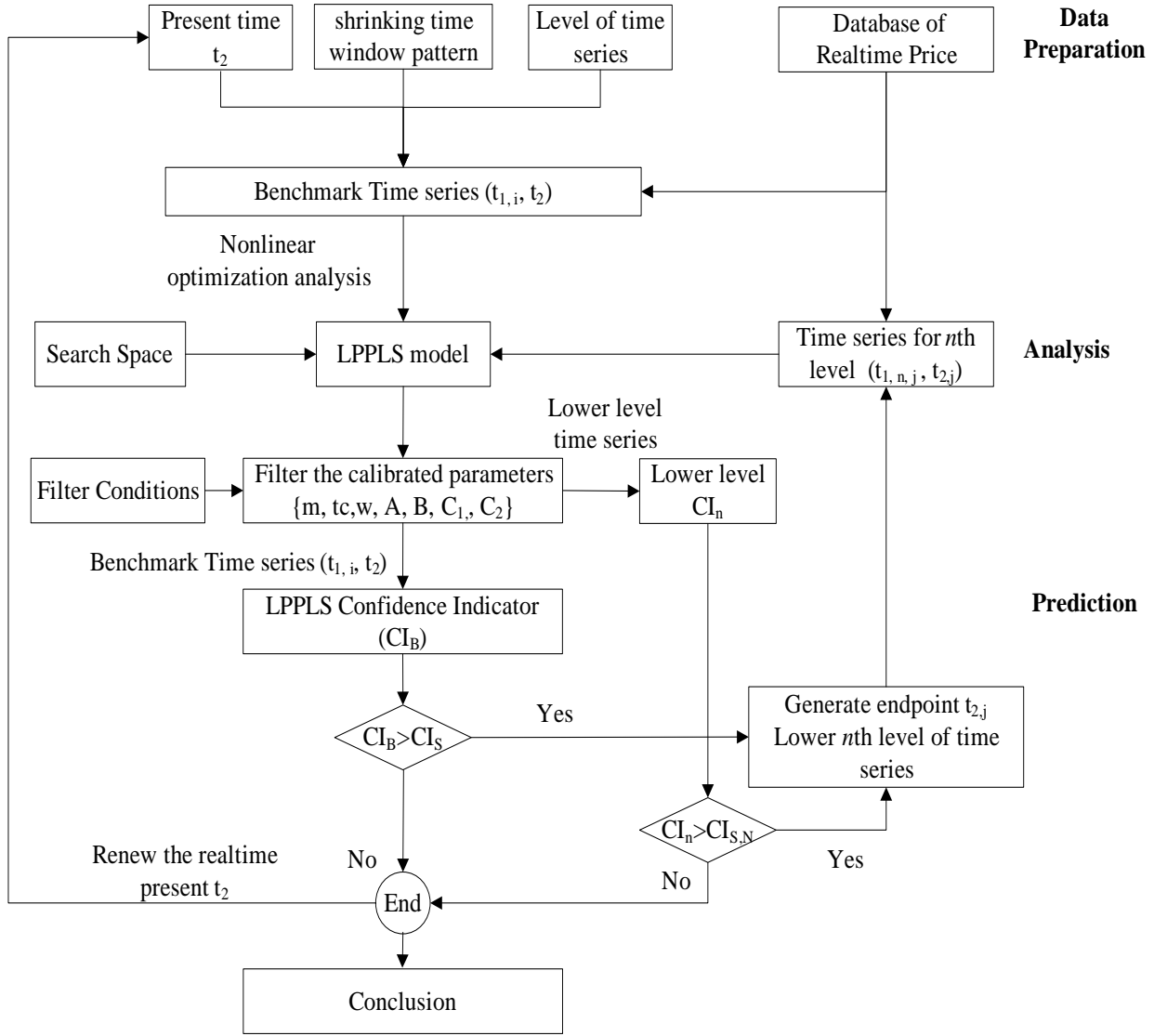

Figure 1. The framework of adaptive multilevel time series detection methodology

In the data preparation phase, we select the present time as the endpoint $t_2$ of the analysis and generate a group of benchmark time series of the cryptocurrency's price based on the specified shrinking time window pattern (e.g., the start time $t_1$ is moved to toward the endpoint $t_2$ decreasing from 650 data points to 30 data points in the step of 5 data points), the collection level of time series (e.g., 1 hour), and the database of real-time price. In the data analysis phase, the nonlinear optimization analysis is performed in the LPPLS model to estimate the best fitted parameters within the specified search space. Based on the specified filter conditions, the calibrated parameters are filtered to select the series of the calibrated parameters which satisfy the specified bubble features. Then, the benchmark LPPLS confidence indicator ($CI_B$) can be calculated as the fraction of fitting windows in which the LPPLS calibrations satisfy the specified filter conditions. In the prediction phrase, the conclusion will be made based on the comparison between the calculated LPPLS confidence indicator and the specified LPPLS confidence indicator ($CI_S$). If the value of $CI_B$ is smaller than the values of $CI_S$, the analysis for bubble detection at the present time $t_2$ will be end and wait to renew the present time $t_2$ to carry out the next loop of the bubble detection. If the value of $CI_B$ exceeds the values of $CI_S$, the number of lower levels of time series will be specified as N (e.g., 2) and the calculation of $CI_n$ will be triggered based on the lower level of time series (e.g., 30min) before and after the present time $t_2$. If the value of $CI_n$ is larger than a specific value $CI_{S,n}$, the next lower level time series (e.g., 15min) will be triggered to calculate $CI_n$ to check if a bubble pattern occurs in the shorter time series. If the $CI_n$ value is smaller than the specified value of $CI_{S,n}$, the prediction will be end and the conclusion can be made based on the above analysis. The benchmark LPPLS confidence indicator $CI_B$ can be used to diagnose bubble in a long-term time scale. The lower level LPPLS confidence indicator $CI_n$ can detect if the bubble pattern occurs in a short-term time scale. The prediction bases on the adaptive multilevel time series detection methodology can effectively diagnose the existence of bubble and accurately predict the bubble crash not only in a long-term time period but also in a short-term time period.

In this study, we adopted two levels of time series: 1 hour and 30 minutes to demonstrate the adaptive multilevel time series detection methodology. The benchmark time series were generated based on the one hour level time series level and shrinking time window by moving the start time $t_1$ toward the endpoint $t_2$ decreasing from 650 data points to 30 data points in the step of 5 data points. For each $t_2$, 125 fitting windows are generated. The value of the specified LPPLS confidence indicator ($CI_S$) is adopted as 0.8% so that the 30 minutes time series level can be triggered as long as there is one bubble pattern existing in the 125 benchmark time series. When $CI_S$ is over 0.8% at a given endpoint time $t_2$ (e.g., 1 am), the lower level $CI_n$ is triggered to calculate before one data point of the $t_2$ (e.g., 0:30 am) and continue to calculate until there is no bubble pattern existing in the lower time series, that is, the value of $CI_n$ becomes zero.

## 3. Empirical analysis

*3.1 Empirical statistics of the Bitcoin crashes*

One of the main empirical and phenomenological characteristics of Bitcoin is the sharp upward and downward price movement with large volatility. Table 1 summarizes the Bitcoin crashes with more than 15% sharp drop within 3 weeks from 9/13/2011 to 4/7/2019. In Table 1, the 51 crashes have been observed just in seven and half years. The crash end day is empirically estimated as the day at the relative low price within 21 days after the Bitcoin price reached to the peak price. The crash duration is calculated as the number of the days from peak to crash end. The crash size is computed as the cumulative return between the peak day and crash end day. It is surprising that the crash size reached to 70.3% and the Bitcoin price dropped dramatically from $229 on 04/9/2013 to 68.1 on 04/16/2013 just in one week. In a recent crash starting on 11/11/2018, the bitcoin price suffered 41.4% loss from $6,357.5 to $3,727.3 within about two weeks.

Figure 1 depicts the Bitcoin crashes with more than 15% sharp drop within 3 weeks from 9/13/2011 to 4/7/2019. As shown in Figure 1 (a), the time gap between crashes can be quite short. Compared to the traditional stock market, the crash duration of Bitcoin presented in Figure 1 (b) can have a shorter value and the price of Bitcoin can lose around 20% value in one day. The number of crashes is significantly different in different years shown in Figure 1 (c). In the year of 2016, 3 crashes occurred. Then the crash number doubled in the each of the following two years, increasing to 6 in 2017 and 11 in 2018. This figure illustrates the fermentation, development, and crash of Bitcoin long timescale bubbles. In the phase of fermentation of Bitcoin bubble, the Bitcoin price is relative stable with small number of crashes. The number of crashes increases as Bitcoin bubble grows. After the long timescale bubbles mature, a succession short timescale crashes can be triggered by any small disturbances due to the systemic instability of the Bitcoin price. In general, the Bitcoin has suffered two long timescale bubbles from 09/13/2011 to 04/07/2019. In the first long timescale bubble, the Bitcoin price increase by 471.4 times within two years, from $2.37 on 11/20/2011 to $1,119.52 on 11/30/2013. The second long timescale bubble occurs between August 2015 to December 2017 and increase by 90.5 times, from $209.7 on 8/24/2015 to $19,187.8 on 12/16/2017. Corresponding to these two long timescale bubbles, the number of Bitcoin crashes climbed to the local peaks at 2014 and 2018, respectively. Figure 1 (d) shows the histogram of Bitcoin crash size. It can be seen that more than half of crash sizes exceeds 25%, indicating the number of Bitcoin crash size has a significant fat tail.

Table 1. Summary of the Bitcoin crashes with more than 15% sharp drop within 3 weeks from 9/13/2011 to 4/7/2019

| Number | Peak Day | Peak Price ($) | Crash End Day | Crash End Price ($) | Crash Duration | Crash Size |
|---|---|---|---|---|---|---|
| 1 | 9/25/2011 | 6.1 | 10/9/2011 | 3.9 | 14 | 35.7% |
| 2 | 10/10/2011 | 4.5 | 10/20/2011 | 2.2 | 10 | 50.3% |
| 3 | 1/10/2012 | 7.1 | 1/28/2012 | 4.9 | 18 | 31.2% |
| 4 | 2/13/2012 | 5.8 | 2/18/2012 | 4.2 | 5 | 26.9% |
| 5 | 8/16/2012 | 13.4 | 8/19/2012 | 8.1 | 3 | 39.9% |
| 6 | 10/10/2012 | 12.2 | 10/26/2012 | 10.0 | 16 | 18.3% |
| 7 | 4/9/2013 | 229.0 | 4/16/2013 | 68.1 | 7 | 70.3% |
| 8 | 4/29/2013 | 143.3 | 5/3/2013 | 98.1 | 4 | 31.6% |
| 9 | 5/29/2013 | 130.4 | 6/14/2013 | 99.0 | 16 | 24.0% |
| 10 | 6/19/2013 | 105.2 | 7/6/2013 | 66.3 | 17 | 37.0% |
| 11 | 10/1/2013 | 127.3 | 10/2/2013 | 103.9 | 1 | 18.4% |
| 12 | 11/18/2013 | 669.0 | 11/19/2013 | 536.0 | 1 | 19.9% |
| 13 | 11/29/2013 | 1132.0 | 12/7/2013 | 693.3 | 8 | 38.8% |
| 14 | 12/10/2013 | 979.2 | 12/18/2013 | 520.0 | 8 | 46.9% |
| 15 | 1/6/2014 | 919.2 | 1/27/2014 | 752.0 | 21 | 18.2% |
| 16 | 2/3/2014 | 808.5 | 2/24/2014 | 535.5 | 21 | 33.8% |
| 17 | 3/5/2014 | 670.0 | 3/23/2014 | 561.0 | 18 | 16.3% |
| 18 | 3/24/2014 | 586.0 | 4/10/2014 | 363.1 | 17 | 38.0% |
| 19 | 4/16/2014 | 530.0 | 5/6/2014 | 428.0 | 20 | 19.2% |

| | | | | | |
|---|---|---|---|---|---|
| 20 | 6/3/2014 | 670.1 | 6/14/2014 | 566.6 | 11 | 15.5% |
| 21 | 8/10/2014 | 591.0 | 8/18/2014 | 475.2 | 8 | 19.6% |
| 22 | 8/29/2014 | 509.3 | 9/19/2014 | 395.9 | 21 | 22.3% |
| 23 | 9/23/2014 | 439.0 | 10/5/2014 | 323.5 | 12 | 26.3% |
| 24 | 11/12/2014 | 426.6 | 11/21/2014 | 351.2 | 9 | 17.7% |
| 25 | 12/7/2014 | 377.3 | 12/18/2014 | 312.7 | 11 | 17.1% |
| 26 | 12/26/2014 | 328.3 | 1/14/2015 | 171.4 | 19 | 47.8% |
| 27 | 3/11/2015 | 296.5 | 3/24/2015 | 245.0 | 13 | 17.4% |
| 28 | 4/5/2015 | 260.5 | 4/14/2015 | 215.8 | 9 | 17.2% |
| 29 | 7/28/2015 | 294.0 | 8/18/2015 | 224.8 | 21 | 23.5% |
| 30 | 8/4/2015 | 285.0 | 8/24/2015 | 209.7 | 20 | 26.4% |
| 31 | 11/4/2015 | 408.0 | 11/11/2015 | 309.9 | 7 | 24.0% |
| 32 | 1/9/2016 | 448.8 | 1/15/2016 | 360.0 | 6 | 19.8% |
| 33 | 6/16/2016 | 766.6 | 6/22/2016 | 601.3 | 6 | 21.6% |
| 34 | 7/17/2016 | 679.5 | 8/2/2016 | 540.0 | 16 | 20.5% |
| 35 | 1/4/2017 | 1114.9 | 1/11/2017 | 778.6 | 7 | 30.2% |
| 36 | 3/3/2017 | 1285.3 | 3/24/2017 | 929.1 | 21 | 27.7% |
| 37 | 6/11/2017 | 2954.2 | 6/15/2017 | 2424.9 | 4 | 17.9% |
| 38 | 7/5/2017 | 2602.9 | 7/16/2017 | 1917.6 | 11 | 26.3% |
| 39 | 9/1/2017 | 4921.7 | 9/14/2017 | 3227.8 | 13 | 34.4% |
| 40 | 11/8/2017 | 7450.3 | 11/12/2017 | 5870.4 | 4 | 21.2% |
| 41 | 12/16/2017 | 19187.8 | 12/30/2017 | 12640.0 | 14 | 34.1% |
| 42 | 1/6/2018 | 17149.7 | 1/18/2018 | 11247.6 | 12 | 34.4% |
| 43 | 1/20/2018 | 12776.0 | 2/5/2018 | 6874.3 | 16 | 46.2% |
| 44 | 3/4/2018 | 11463.3 | 3/17/2018 | 7860.8 | 13 | 31.4% |
| 45 | 3/23/2018 | 8920.8 | 4/6/2018 | 6618.3 | 14 | 25.8% |
| 46 | 5/5/2018 | 9823.3 | 5/26/2018 | 7336.0 | 21 | 25.3% |
| 47 | 6/7/2018 | 7689.3 | 6/28/2018 | 5848.3 | 21 | 23.9% |
| 48 | 7/24/2018 | 8403.8 | 8/10/2018 | 6140.0 | 17 | 26.9% |
| 49 | 9/4/2018 | 7361.0 | 9/8/2018 | 6178.3 | 4 | 16.1% |
| 50 | 11/11/2018 | 6357.5 | 11/26/2018 | 3727.3 | 15 | 41.4% |
| 51 | 11/29/2018 | 4249.8 | 12/15/2018 | 3179.5 | 16 | 25.2% |

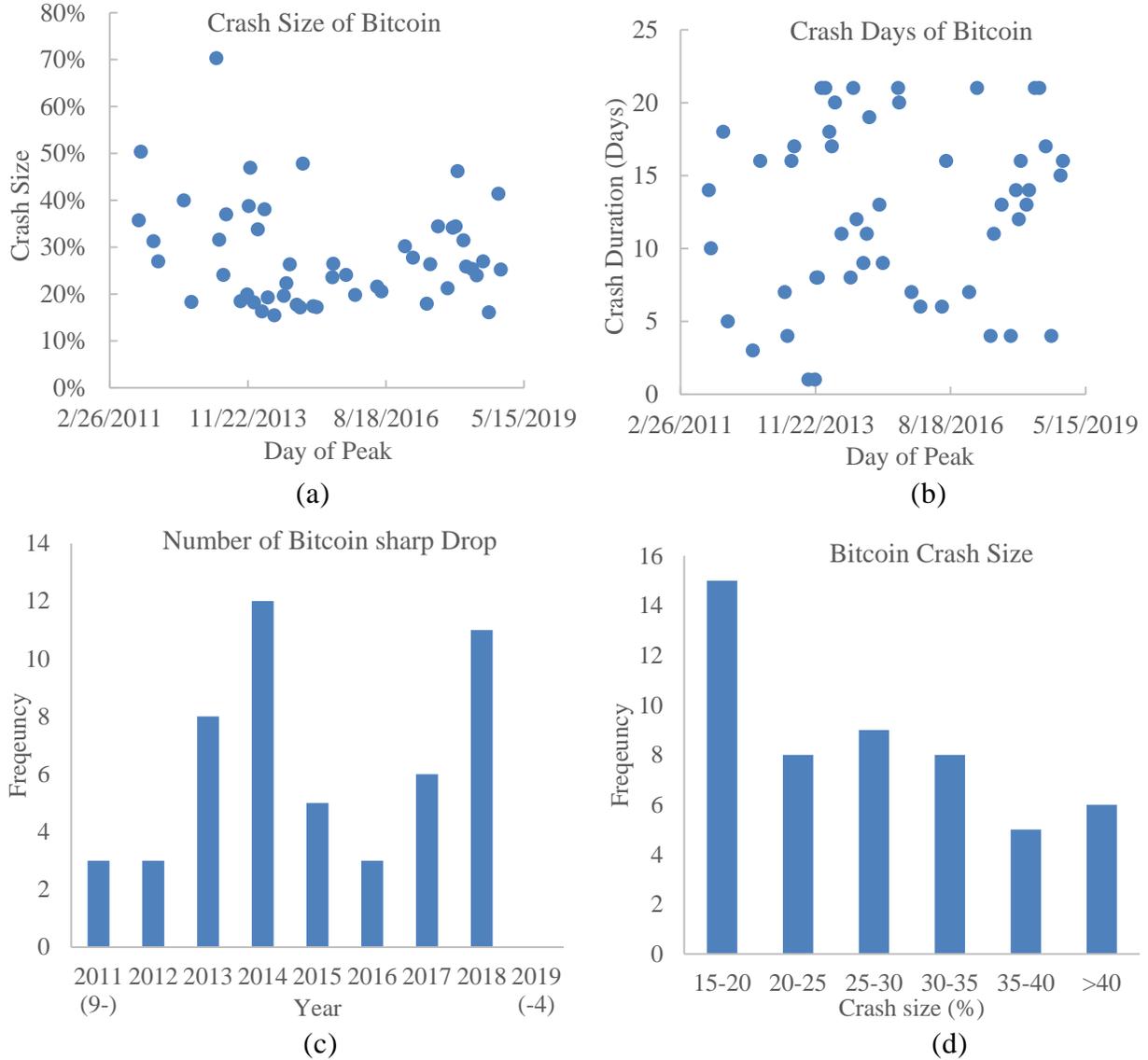

Figure 1. Statistics of the Bitcoin crashes indicate more than 15% sharp drop within 3 weeks from 9/13/2011 to 4/7/2019, with (a) Crash size, (b) Crash duration, (c) Frequency of Bitcoin sharp drop, and (d) Frequency of Bitcoin crash size.

*3.2 LPPLS Bubble Identification Based on Daily Data of Bitcoin Price*

In this section, we use the LPPLS confidence indicator as a diagnostic tool for identifying both the positive and negative bubbles based on the daily data of Bitcoin price from 1/11/2017 to 4/11/2019. The positive bubbles are associated with upwardly accelerating price increase and are susceptible to regime changes in the form of crashes or volatile sideway plateaus, while the negative bubbles are associated with downwardly accelerating price decrease and are susceptible to regime changes in the form of valleys or volatile sideway plateaus. The endpoint time $t_2$ is moved from 1/11/2017 to 4/11/2019 in the step of one day to generate 821 number of the $t_2$. For a specified $t_2$, the shrinking time windows are generated by moving the start time $t_1$ toward $t_2$ decreasing from 650 days to 30 days in the step of 5 days. Thus, there are 125 fitting time series for each $t_2$.

Figure 2 shows the LPPLS confidence indicator for positive bubbles in red and negative bubble in green along with the Bitcoin price in blue from 1/11/2017 to 4/11/2019. The confidence level of the observed LPPLS bubble pattern can be intuitively reflected in Figure 2. If the bubble pattern existed in most of the fitting time windows, the value of LPPLS confidence indicator can be up to one indicating the indicator is no sensitivity to the choice of the starting time $t_1$. If the indicator has a value close to zero, the results need careful consideration due to the over-fitting risk. In Figure 2, there are five obvious clusters of positive bubbles and several clusters of negative bubbles. The first cluster of positive bubble takes place between March 1, 2017 and March 7, 2017. Bitcoin price goes up to a peak value of $1,285.3 on March 3, 2017, and then goes down to a valley of $929.1 on March 24, 2017. This valley period was also confirmed by a cluster of negative bubbles. The second cluster of positive bubbles is observed from May 17, 2017 to July 4, 2017, and the LPPLS confidence indicator reaches to the peak value of 0.072 on June 11, 2017, indicating the bubble pattern existed in nine fitting time windows.

The prediction is confirmed by the fact that the Bitcoin price dropped dramatically from $2,954.2 on June 11, 2017 to $1,917.6 on July 16, 2017, losing 35.1% value within 35 days. The valley on July 16, 2017 is predicted by a cluster of negative bubbles from July 14, 2017 to July 19, 2017. The third obvious cluster of positive bubbles occurs from August 22, 2017 to September 3, 2017, corresponding to the fact that the Bitcoin price surges to $4,921.7 on September 1, 2017, and then decreases by 34.4% to $3,227.8 after 13 days. The fourth obvious cluster of positive bubbles happens between November 3, 2017 to November 9, 2017 and the Bitcoin price decreases by 21.2% of value from $7,450.3 on November 8, 2017 to $5,870.4 on November 12, 2017. The last cluster of positive bubble occurs after April 1, 2019 to the end time of this analysis and the Bitcoin price rockets to $5,324.6 on April 10, 2019, increasing by 28.1% within 10 day. A negative bubble cluster is observed from February 1, 2018 to February 7, 2018 and the Bitcoin price falls steeply to $6,955.3 on February 5, 2018 from $17,527 on January 6, 2018, losing 60.3% value within one month. From March 5, 2018 to April 6, 2018, the Bitcoin price declines by 42.7% from $11,573.3 to $6,636.3 and the lowest valley falls into the range of the negative cluster between April 5, 2018 and April 9, 2018. The negative cluster from June 16, 2018 to June 29, 2018 successfully forecasts the lowest valley on June 28, 2018 and Bitcoin price decreases by 39.4% from May 3, 2018.

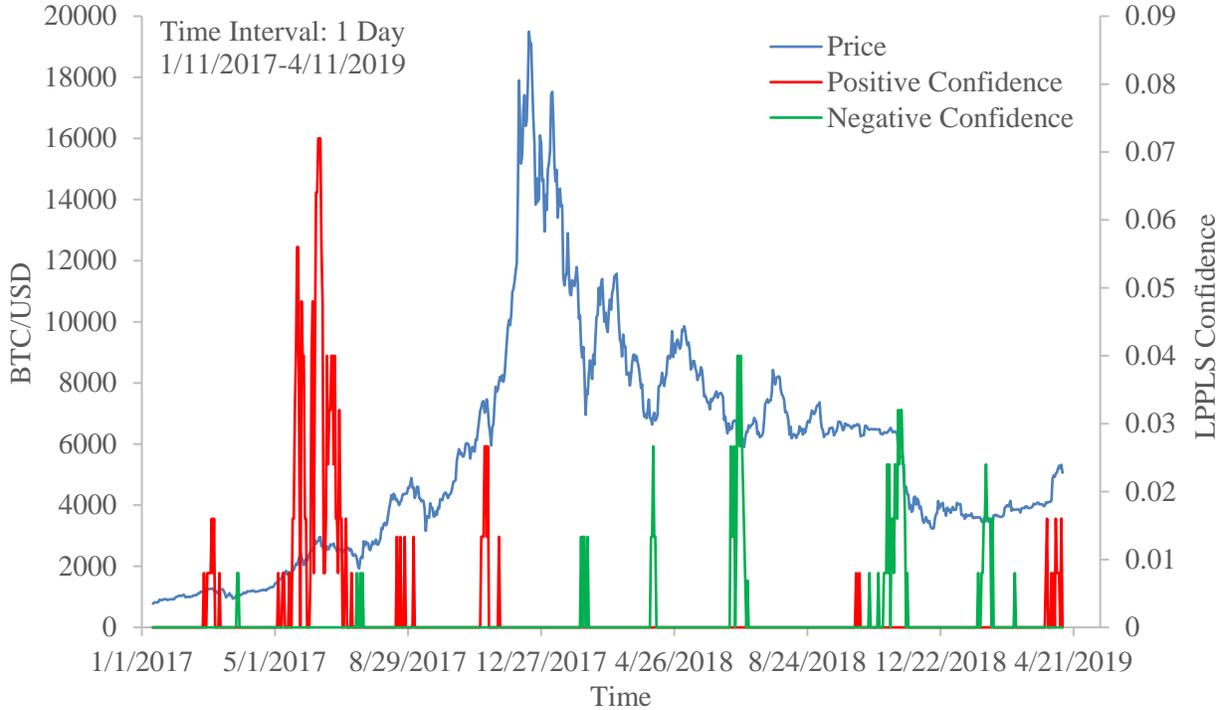

Figure 2. Daily LPPLS confidence indicator is shown for positive bubbles in red and negative bubble in green (right scale) along with the Bitcoin price in blue (left scale) from 1/11/2017 to 4/11/2019.

Although the LPPLS confidence indicator based on the daily data of Bitcoin price has successfully predicted some positive bubbles and negative bubbles of the Bitcoin price, it does not provide effective warnings for detecting the bubbles when the Bitcoin price suffers from a large fluctuation in a short time period, especially for positive bubbles. Between December 2017 and August 2018, the Bitcoin price has experienced a roller coaster ride as shown in Figure 2. However, the LPPLS confidence indicator fails to provide any useful diagnosis of the existence of bubbles because the daily data of Bitcoin price do not provide sufficient information on features of the dramatic price changes for the LPPLS model to determine the bubble pattern effectively.

*3.3 LPPLS Bubble Identification Using Adaptive Multilevel Time Series Detection Methodology*

Based on the adaptive multilevel time series detection methodology introduced in Section 2.3, we have diagnosed the existence of bubbles and monitored the development of bubbles in Bitcoin price sequence between 10/28/2017 and 6/30/2018. In Section 3.1, we can see that Bitcoin price is characterized by the frequently sharp upward and downward swing with large volatility and the time period of bubble fermentation, development, mature and bursts is significantly short in comparison to the traditional stock market. Even the time gap between crashes can be very short. The daily data used in the analysis of the traditional stock market cannot capture the changing features of Bitcoin price and provide sufficient data points to predict the Bitcoin price movement. In order to capture the features of price fluctuation accurately, the time interval of the price data must be more refined. In this study, we adopt two levels of time series, 1 hour and 30 minutes, to demonstrate and verify the adaptive multilevel time series detection methodology. For each day, we can collect 24 data points for 1 hour time interval and 48 data points for 30 minutes time interval as the Crypto market usually operates 24/7. For a specified endpoint $t_2$, the benchmark time series based on the one-hour level of time interval are generated by moving start time $t_1$ toward the $t_2$ increasing from 30 hours to 650 hours in the step of 5 hours. The time scale of one-hour time interval ranges approximately from one day to one month. If a bubble pattern is detected at the specified $t_2$ based on the one hour level of time series, the 30 minutes level of time series will be triggered to generate for the endpoint time which is 30 minutes before the current $t_2$ by increasing shrinking time window length from 30 data points to 650 data points in the step of 5 data points in the 30 minutes level of time series. The time scale of 30 minutes time interval goes approximately from half day to two weeks. The LPPLS confidence indicator for 30 minutes level of time series will continue to calculate for the following endpoint time until there is no bubble pattern existing at the following endpoint. The LPPLS confidence indicator based on the one hour time interval can provide a relative long-term diagnosis of bubble status, while the LPPLS confidence indicator based on the 30 minutes time interval can detect the bubble conditions in a relative short-term time and verify the bubble status obtained from the analysis based on the one hour time interval. The adaptive multilevel time series detection methodology can provide a highly sensitive real-time predication of the bubble status.

Figure 3 shows the LPPLS confidence indicator for positive bubbles in red and negative bubble in green along the Bitcoin price in blue from 10/28/2017 21:00 to 6/30/2018 0:00 using the two levels of the adaptive multilevel time series detection methodology. From Figure 3, we can see that the dramatic changes of Bitcoin price and rapid switches of bubble status which causes the failure of diagnosis of the existence of bubbles based on the daily data of Bitcoin price discuss in Section 3.2 can be predicted successfully using the two levels of the adaptive multilevel time series detection methodology. In Figure 3, the LPPLS confidence indicator for the positive bubble reaches the maximum value of 68.8% on December 6, 2017 20:00, indicating the bubble pattern can be found in 86 out of 125 fitting windows and Bitcoin price has already reached the systemic instability so that any small disturbances can cause the mature bubble to burst.

Thirteen obvious clusters of positive bubbles and negative bubbles can be observed in Figure 3. Table 2 summarizes the details of the clusters of positive bubbles and negative bubbles and related information of the actual peaks and valleys. In Table 2, it can be noted that the actual peak time and actual valley time always falls into the range of the clusters of positive bubble and the clusters of negative bubbles, respectively, indicating that the LPPLS confidence indicator based on the adaptive multilevel time series detection methodology have an outstanding performance to diagnose the existence of bubble and accurately predict the bubble crash. For example, an obvious cluster of positive bubble formed from 10/31/2017 22:00 to 11/7/2017 7:00 and reached a peak value of 35.2% on 11/2/2017 11:00, followed by a negative bubble cluster between 11/10/2017 14:30 and 11/13/2017 9:00 with the peak value of 24.0% on 11/13/2017 1:30. This diagnosis of the bubble status is verified by the price movement of the Bitcoin which jumps to the peak value of $7,571 on 11/5/2017 13:00 and then decreased by 32.8% to $5,700 on 11/12/2017 22:00. In addition, the LPPLS confidence indicator based on the adaptive multilevel time series detection methodology can monitor the development and crash of bubble even if the bubble exists in a short time. For example, the Bitcoin price dramatically increased from $12,953

on 01/12/2018 0:00 to $14,492 on 01/13/2018 13:30 in less than two days. A cluster of positive bubble formed from 01/13/2018 5:00 to 01/13/2018 22:30, which covered the actual peak time.

It can be noted that the clusters of the positive bubbles sometimes have overlapped with the clusters of the negative bubbles in Figure 3. That is because the Bitcoin price may be suffering a valley in a short-term time scale given a specified time while it may be also approaching a peak in a long-term time scale, or the Bitcoin price may have a peak value in a short-term time scale, but at the same time, it may reach a lowest value in a long-term time scale.

In order to measure the bubble status with different time scale, we partition the window sizes to calculate the LPPLS confidence indicator into two classes: short-term time scale and long-term time scale. In the short-term time scale, the start time $t_1$ is moved to toward the endpoint $t_2$ increasing from 30 data points to 200 data points in the step of 5 data points, so that 35 fitting time windows are generated to diagnose the short-term bubble status. In the long-term time scale, the length of the shrinking time windows increases from 205 data points to 650 data points in the step of 5 data points to generate 90 fitting time windows to detect the long-term bubble status. For the one hour benchmark time series, the short-term time scale goes approximately from one day to one week and the long-term time scale ranges from one week to one month. Figure 4 and Figure 5 shows the LPPLS confidence indicator for positive bubbles in red and negative bubble in green along with the Bitcoin price in blue from 10/28/2017 to 6/30/2018 in the short-term time scale and long-term time scale, respectively. It can be seen that the cluster number of the positive and negative bubbles in short-term time scale is much more than the one in the long-term time scale. The short-term LPPLS confidence indicator is highly sensitive to the extreme price fluctuations of Bitcoin price and provides the useful insights into the bubble status in a shorter time scale - on a day to week scale. In comparison, the long-term LPPLS confidence indicator has a stable performance and monitor the bubble status in a longer time scale - on a week to month scale. We can effectively diagnose the existence of bubble at a specified time not only in a short-term time scale but also in a long-term time scale by accounting for both the short-term and long-term LPPLS confidence indicators.

To improve the diagnosis performance of the adaptive multilevel time series detection methodology, we recommend that multiple levels of time series can be adopted in the analysis, for example, 1 day, 6 hours, 3 hours, 1 hour, 30 minutes, 15 minutes, and 5 minutes. Different levels of the time series can provide bubble detections with different time scales.

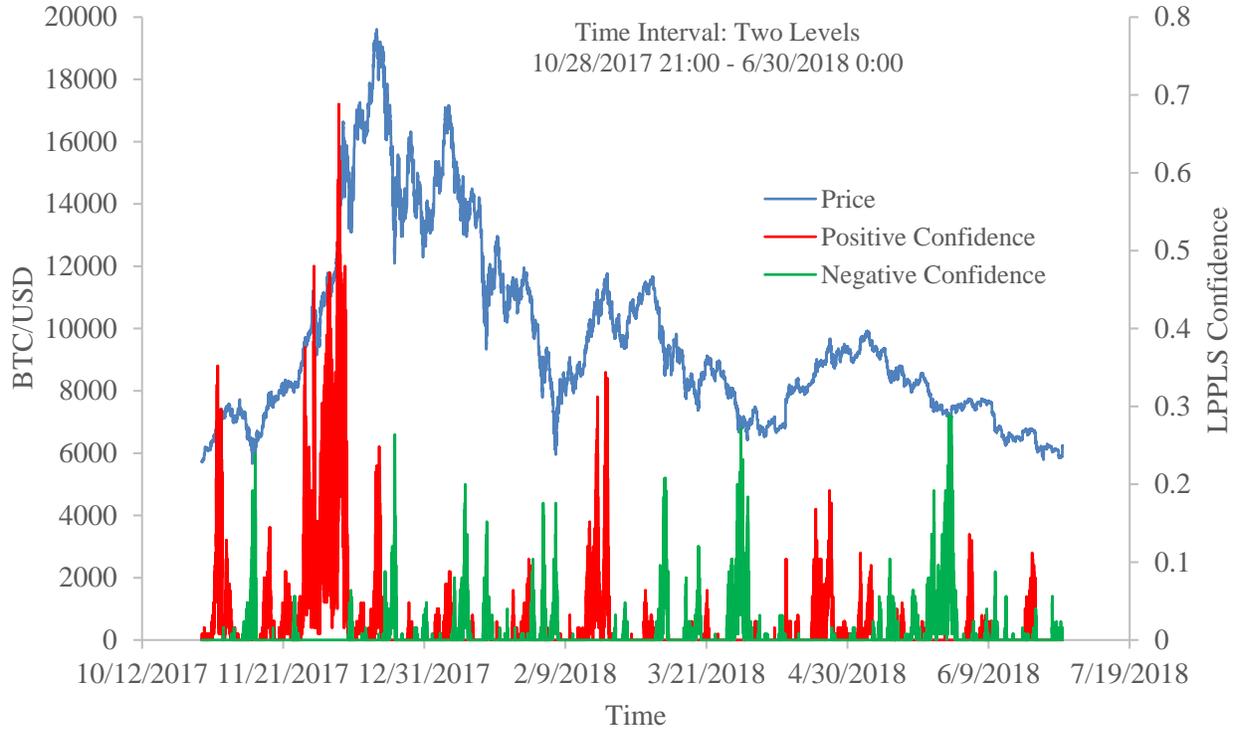

Figure 3. LPPLS confidence indicator for positive bubbles is shown in red and negative bubble in green (right scale) along with the Bitcoin price in blue (left scale) from 10/28/2017 to 06/30/2018

Table 2. Summary of the Bitcoin crashes detected by LPPLS confidence indicator from 10/28/2017 21:00 to 6/30/2018 0:00 using the two levels of the adaptive multilevel time series detection methodology

| Crash No | Cluster of Positive bubble | | Cluster of Negative bubble | |
|---|---|---|---|---|
| | Start Time | End Time | Start Time | End Time |
| 1 | 10/31/2017 22:00 | 11/7/2017 7:00 | 11/10/2017 14:30 | 11/13/2017 9:00 |
| 2 | 11/25/2017 6:00 | 12/9/2017 11:00 | 12/9/2017 17:00 | 12/10/2017 16:30 |
| 3 | 12/12/2017 0:00 | 12/19/2017 10:00 | 12/19/2017 17:00 | 12/23/2017 1:30 |
| 4 | 12/26/2017 8:00 | 12/27/2017 13:30 | 12/28/2017 6:00 | 12/31/2017 18:30 |
| 5 | 1/5/2018 11:00 | 1/7/2018 23:30 | 1/8/2018 6:30 | 1/12/2018 13:30 |
| 6 | 1/13/2018 5:00 | 1/13/2018 22:30 | 1/16/2018 13:30 | 1/19/2018 10:00 |
| 7 | 1/27/2018 12:00 | 1/30/2018 12:30 | 2/2/2018 10:00 | 2/6/2018 18:00 |
| 8 | 2/14/2018 17:00 | 2/21/2018 12:30 | 2/25/2018 8:00 | 2/26/2018 12:00 |
| 9 | 3/3/2018 6:30 | 3/6/2018 7:00 | 3/6/2018 18:00 | 3/10/2018 5:30 |
| 10 | 3/20/2018 16:30 | 3/21/2018 16:00 | 3/26/2018 22:00 | 4/1/2018 22:00 |
| 11 | 4/20/2018 0:00 | 5/7/2018 3:00 | 5/22/2018 10:00 | 5/31/2018 3:00 |
| 12 | 6/6/2018 20:30 | 6/9/2018 22:00 | 6/10/2018 17:00 | 6/11/2018 12:30 |
| 13 | 6/19/2018 8:00 | 6/22/2018 8:00 | 6/26/2018 23:00 | 6/29/2018 20:00 |

Table 2 (continued)

| No | Actual Peak Time | Actual Peak Price | Actual Valley Time | Actual Valley Price | Crash Size |
|---|---|---|---|---|---|
| 1 | 11/5/2017 13:00 | 7571 | 11/12/2017 22:00 | 5700 | 32.80% |
| 2 | 12/8/2017 1:00 | 16626 | 12/10/2017 3:00 | 13110 | 21.10% |
| 3 | 12/17/2017 12:00 | 19600 | 12/22/2017 14:00 | 12092 | 38.30% |
| 4 | 12/27/2017 5:00 | 16320 | 12/30/2017 16:00 | 12300 | 24.60% |
| 5 | 1/6/2018 23:00 | 17166 | 1/12/2018 0:00 | 12953 | 24.50% |
| 6 | 1/13/2018 13:30 | 14492 | 1/17/2018 14:00 | 9500 | 34.40% |
| 7 | 1/28/2018 6:00 | 11948 | 2/6/2018 4:00 | 6100 | 48.90% |
| 8 | 2/20/2018 17:00 | 11670 | 2/25/2018 18:00 | 9354 | 19.80% |
| 9 | 3/5/2018 18:00 | 11667 | 3/9/2018 9:00 | 8558 | 26.70% |
| 10 | 3/21/2018 4:00 | 9122 | 4/1/2018 15:00 | 6429 | 29.50% |
| 11 | 5/5/2018 10:00 | 9908 | 5/29/2018 2:00 | 7089 | 28.50% |
| 12 | 6/7/2018 1:00 | 7738 | 6/10/2018 22:00 | 6694 | 13.50% |
| 13 | 6/21/2018 2:00 | 6775 | 6/29/2018 13:00 | 5861 | 13.50% |

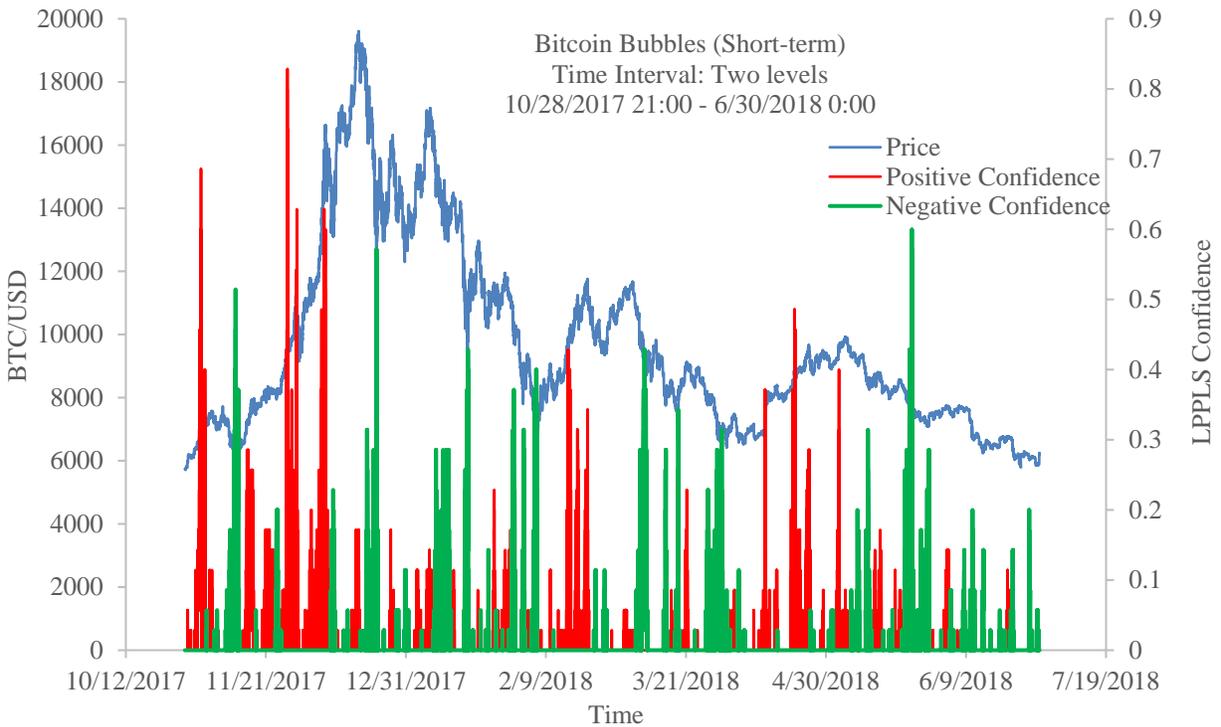

Figure 4. Short-term LPPLS confidence indicator for positive bubbles is shown in red and negative bubble in green (right scale) along with the Bitcoin price in blue (left scale) from 10/28/2017 to 06/30/2018

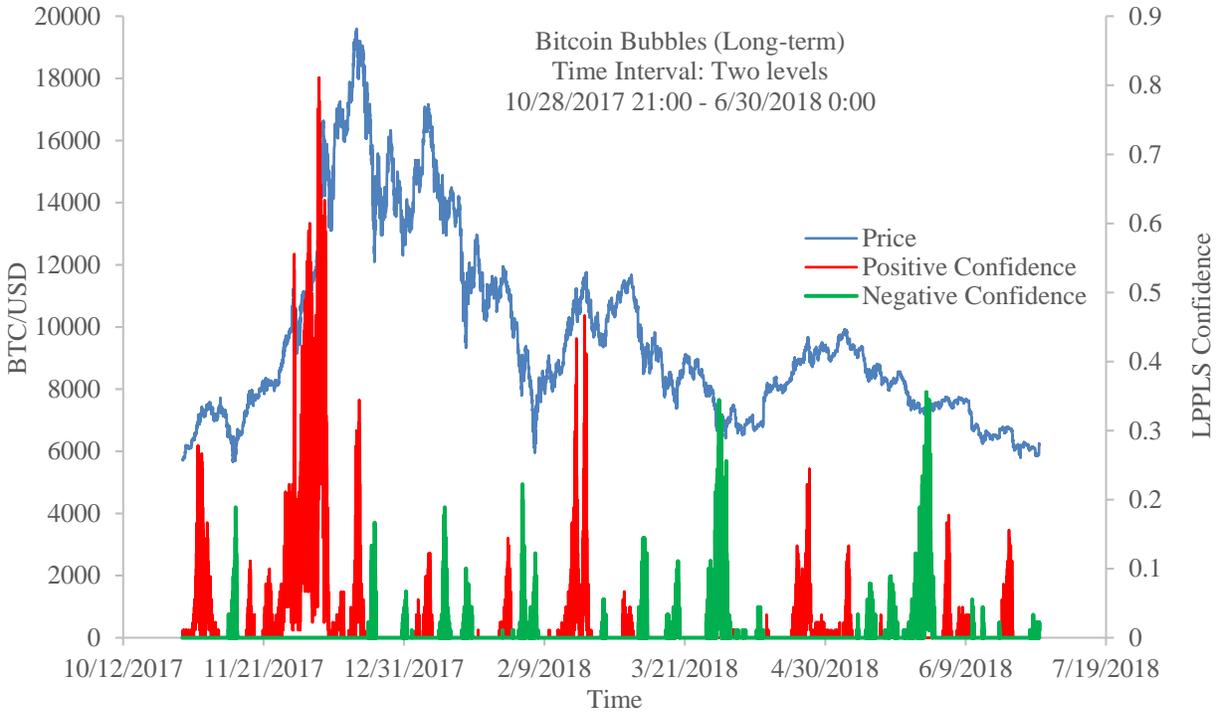

Figure 5. Long-term LPPLS confidence indicator for positive bubbles is shown in red and negative bubble in green (right scale) along with the Bitcoin price in blue (left scale) from 10/28/2017 to 06/30/2018

## 4. Conclusions

In this study, we summarize the Bitcoin historical crashes at the magnitude of more than 15% sharp drop within 3 weeks from 09/13/2011 to 04/07/2019 and observe a total of 51 crashes for this very recent 7.5-year period. Based on the statistical analysis of the Bitcoin price, its crash duration and the time gap between crashes can be very short while more than half of the crashes saw over 25% of a price drop. The number of crashes increases as the Bitcoin long timescale bubble grows, and a succession of short timescale crashes can be triggered as the long timescale bubble matures.

Based on the LPPLS model, we are able to identify both the positive and negative bubbles using the daily data of Bitcoin price from 01/11/2017 to 04/11/2019. Although the LPPLS confidence indicator based on the daily data of Bitcoin price has provided useful predictions of some positive as well as negative bubbles, it fails to provide effective warnings for detecting the bubbles when the Bitcoin price suffers from a large fluctuation in a short time period, especially for positive bubbles.

In order to diagnose the existence of bubble and accurately predict the bubble crash, we propose an adaptive multilevel time series detection methodology. Based on the adaptive multilevel time series detection methodology and finer (than daily) Bitcoin price data, we adopt two levels of time series, 1 hour and 30 minutes, to diagnose the existence of bubbles and monitor the development of the bubbles in the Bitcoin price sequence between 10/28/2017 and 06/30/2018. Our analysis shows that the LPPLS confidence indicator based on the adaptive multilevel time series detection methodology has an outstanding performance in bubble detection and crash forecast, even if the bubble exists in a short time period. We find that the short-term LPPLS confidence indicator is highly sensitive to extreme price fluctuations and can therefore provide useful insights into the bubble status on a shorter time scale - on a day to week scale. We also observe that the long-term LPPLS confidence indicator features a stable performance that can effectively monitor the bubble status on a longer time scale - on a week to month scale. In summary, the newly proposed adaptive multilevel time series detection methodology can provide not only real-time detection of bubbles but also early warning of bubble crashes. Furthermore, our methodology is applicable to not only the cryptocurrency market but also other financial markets in general.

**Acknowledgment**

The authors would like to thank the Stony Brook Research Computing and Cyberinfrastructure, and the Institute for Advanced Computational Science at Stony Brook University for access to the high-performance SeaWulf computing system, which was made possible by a $1.4M National Science Foundation grant (#1531492).

**References**

[1] I. Allison, Nick Szabo: If banks want benefits of blockchains they must go permissionless, https://www.ibtimes.co.uk/nick-szabo-if-banks-want-benefits-blockchains-they-must-go-permissionless-1518874, 2019 (accessed 4 April 2019).
[2] S. Nakamoto, Bitcoin: A peer-to-peer electronic cash system, (2008).
[3] Statista, Number of Blockchain wallet users worldwide from 1st quarter 2015 to 4th quarter 2018, https://www.statista.com/statistics/647374/worldwide-blockchain-wallet-users/, 2019 (accessed 4 April 2019).
[4] A. Lielacher, How Many People Use Bitcoin in 2019?, https://www.bitcoinmarketjournal.com/how-many-people-use-bitcoin/, 2019 (accessed 4 April 2019).
[5] B.P. Hanley, The false premises and promises of Bitcoin, arXiv preprint arXiv:1312.2048, (2013).
[6] D. Yermack, Is Bitcoin a real currency? An economic appraisal, in: D.L.K. Chuen (Ed.) Handbook of digital currency, Academic Press, 2015, pp. 31-43.
[7] N. Popper, Digital gold: The untold story of Bitcoin, Penguin, London, UK, 2015.
[8] R. Browne, Nobel prize-winning economist Robert Shiller thinks bitcoin is an "interesting experiment", https://www.cnbc.com/2018/01/26/robert-shiller-says-bitcoin-is-aninteresting-experiment.html, 2018 (accessed 7 April 2019).
[9] L. Kristoufek, BitCoin meets Google Trends and Wikipedia: Quantifying the relationship between phenomena of the Internet era, Sci. Rep., 3 (2013) 3415.
[10] D. Garcia, C.J. Tessone, P. Mavrodiev, N. Perony, The digital traces of bubbles: feedback cycles between socio-economic signals in the Bitcoin economy, J. Royal Soc. Interface, 11 (2014) 20140623.
[11] F. Glaser, K. Zimmermann, M. Haferkorn, M.C. Weber, M. Siering, Bitcoin-asset or currency? revealing users' hidden intentions, in: Twenty Second European Conference on Information Systems, Tel Aviv, Israel, 2014.

[12] J. Donier, J.-P. Bouchaud, Why do markets crash? Bitcoin data offers unprecedented insights, PLoS One, 10 (2015) e0139356.
[13] E. Bouri, P. Molnár, G. Azzi, D. Roubaud, L.I. Hagfors, On the hedge and safe haven properties of Bitcoin: Is it really more than a diversifier?, Financ. Res. Lett., 20 (2017) 192-198.
[14] M. Balcilar, E. Bouri, R. Gupta, D. Roubaud, Can volume predict Bitcoin returns and volatility? A quantiles-based approach, Econ. Model., 64 (2017) 74-81.
[15] A.F. Bariviera, The inefficiency of Bitcoin revisited: A dynamic approach, Econ. Lett., 161 (2017) 1-4.
[16] S. Begušić, Z. Kostanjčar, H.E. Stanley, B. Podobnik, Scaling properties of extreme price fluctuations in Bitcoin markets, Physica A, 510 (2018) 400-406.
[17] S. McNally, J. Roche, S. Caton, Predicting the price of Bitcoin using Machine Learning, in: 26th Euromicro International Conference on Parallel, Distributed and Network-based Processing, IEEE, Cambridge, UK, 2018, pp. 339-343.
[18] S. Wheatley, D. Sornette, T. Huber, M. Reppen, R.N. Gantner, Are Bitcoin bubbles predictable? Combining a generalized Metcalfe's Law and the Log-Periodic Power Law Singularity model, R. Soc. open sci., 6 (2019) 180538.
[19] J.-C. Gerlach, G. Demos, D. Sornette, Dissection of Bitcoin's Multiscale Bubble History from January 2012 to February 2018, arXiv preprint arXiv:1804.06261, (2018).
[20] K. Ide, D. Sornette, Oscillatory finite-time singularities in finance, population and rupture, Physica A, 307 (2002) 63-106.
[21] V. Filimonov, D. Sornette, A stable and robust calibration scheme of the log-periodic power law model, Physica A, 392 (2013) 3698-3707.
[22] L. Lin, R.E. Ren, D. Sornette, The volatility-confined LPPL model: A consistent model of 'explosive' financial bubbles with mean-reverting residuals, Int. Rev. Finan. Anal., 33 (2014) 210-225.
[23] D. Sornette, G. Demos, Q. Zhang, P. Cauwels, V. Filimonov, Q. Zhang, Real-time prediction and post-mortem analysis of the Shanghai 2015 stock market bubble and crash, J. Invest. Strategies, 4 (2015) 77–95.
[24] V. Filimonov, G. Demos, D. Sornette, Modified profile likelihood inference and interval forecast of the burst of financial bubbles, Quant. Finance, 17 (2017) 1167-1186.
[25] D. Sornette, Discrete-scale invariance and complex dimensions, Phys. Rep., 297 (1998) 239-270.
[26] W. Yan, Identification and Forecasts of Financial Bubbles, in: Department of Management, Technology and Economics ETH Zurich, 2011.
[27] D. Sornette, A. Johansen, J.-P. Bouchaud, Stock market crashes, precursors and replicas, J. Phys. I, 6 (1996) 167-175.
[28] A. Johansen, O. Ledoit, D. Sornette, Crashes as critical points, Int. J. Theor. Appl. Finance, 3 (2000) 219-255.
[29] D. Sornette, A. Johansen, Large financial crashes, Physica A, 245 (1997) 411-422.
[30] D. Sornette, Critical market crashes, Phys. Rep., 378 (2003) 1-98.
[31] N. Hansen, A. Ostermeier, A. Gawelczyk, On the Adaptation of Arbitrary Normal Mutation Distributions in Evolution Strategies: The Generating Set Adaptation, in: L. Eshelman (Ed.) the Sixth International Conference on Genetic Algorithms, Morgan Kaufmann, San Francisco, CA, 1995, pp. 57-64.
[32] Z.-Q. Jiang, W.-X. Zhou, D. Sornette, R. Woodard, K. Bastiaensen, P. Cauwels, Bubble diagnosis and prediction of the 2005–2007 and 2008–2009 Chinese stock market bubbles, J. Econ. Behav. Organ., 74 (2010) 149-162.

[33] H.-C. Bothmer, C. Meister, Predicting critical crashes? A new restriction for the free variables, Physica A, 320 (2003) 539-547.

[34] Y. Huang, A. Johansen, M. Lee, H. Saleur, D. Sornette, Artifactual log-periodicity in finite size data: Relevance for earthquake aftershocks, J Geophys Res Solid Earth, 105 (2000) 25451-25471.

[35] D. Sornette, W.X. Zhou, The US 2000-2002 market descent: How much longer and deeper?, Quant. Finance, 2 (2002) 468-481.